\documentclass[letterpaper, 10 pt, conference]{ieeeconf}  

\IEEEoverridecommandlockouts                              
\usepackage[ruled,linesnumbered]{algorithm2e}
\DeclareMathAlphabet{\mathcal}{OMS}{cmsy}{m}{n}
\usepackage{amssymb,amsmath,bm}
\usepackage{mathtools}

\usepackage{hyperref,url}
\usepackage{comment}

\usepackage{wasysym}

\usepackage{soul}

\usepackage[dvipsnames]{xcolor}
\usepackage{xspace}

\begin{document}

\title{\LARGE \bf
System-Wide Emergency Policy \\ for Transitioning from Main to Secondary Fuel}

\author{Laurent Pagnier, Criston Hyett, Robert Ferrando, Igal Goldshtein, \\Jean Alisse, Lilah Saban, and Michael Chertkov
\thanks{L. Pagnier, C. Hyett, R. Ferrando, and M. Chertkov are with the Graduate Interdisciplinary Program in Applied Mathematics and Department of Mathematics, University of Arizona, Tucson, AZ 85721, USA. {\tt\small \{laurentpagnier, cmhyett, rferrando, chertkov\}@arizona.edu.}}%
\thanks{J. Alisse, I. Goldshtein, and L. Saban are with Noga, the Israel Independent System Operator, Haifa, Israel.  {\tt\small \{Jean.Alisse, Igal.Goldshtein, Lilah.Saban\}@noga-iso.co.il.}}%
}

\maketitle
\pagestyle{empty}

\begin{abstract}

Faced with the complexities of managing natural gas-dependent power system amid the surge of renewable integration and load unpredictability, this study explores strategies for navigating emergency transitions to costlier secondary fuels. Our aim is to develop decision-support tools for operators during such exigencies. We approach the problem through a Markov Decision Process (MDP) framework, accounting for multiple uncertainties. These include the potential for dual-fuel generator failures and operator response during high-pressure situations. Additionally, we consider the finite reserves of primary fuel, governed by gas-flow partial differential equations (PDEs) and constrained by nodal pressure. Other factors include the variability in power forecasts due to renewable generation and the economic impact of compulsory load shedding. For tractability, we address the MDP in a simplified context, replacing it by Markov Processes evaluated against a selection of policies and scenarios for comparison. Our study considers two models for the natural gas system: an over-simplified model tracking linepack and a more nuanced model that accounts for gas flow network heterogeneity. The efficacy of our methods is demonstrated using a realistic model replicating Israel's power-gas infrastructure. 
\end{abstract}

\section{Introduction}

The coupling between gas and power grids, particularly from the perspective of power system operators, has been well-explored in the literature. For instance, \cite{chertkov_cascading_2015} examines the effect of increased wind penetration as a renewable energy source on the connected natural gas system. The dependency of unit commitment in power systems on gas infrastructure is addressed in \cite{zlotnik_coordinated_2016}, which explores coordination methods between the two systems. A simpler, steady-state model of the gas system is used in \cite{byeon2020awareness} to develop a gas-aware unit commitment problem. In contrast, dynamic models are reintroduced in \cite{Bayani_2022}, where a rank minimization approach is employed to solve the gas-aware look-ahead commitment problem more tractably. Finally, \cite{roald2020uncertainty} acknowledges the exchange of uncertainty between power and gas systems via gas-fired generators and proposes methods for managing this uncertainty.

It is important to note, however, that much of this research focuses on standard operational conditions. The critical issue of gas-grid interdependency during emergency scenarios remains largely unexplored.

In addition, the availability of alternative fuel sources for gas-fired generators is often overlooked. Nevertheless, the advantages of maintaining a significant dual-fuel capability within the generation fleet --particularly in the face of potential supply disruptions -- are well recognized. According to \cite{eia2023}, approximately 13\% of U.S. electricity generation capacity has the flexibility to switch between natural gas and oil. However, the specific challenges regarding timing, reliability, and the complexities involved in transitioning from primary to secondary fuel sources at scale remain largely unexplored.

Notably, this problem is of a special significance for energy management in quite a number of relatively small and partially isolated countries or regions, including Ireland \cite{leahy2012cost}, South Korea \cite{kim2015estimation}, Texas, US \cite{abu-khalifa2023}, and Israel, to name a few. In this article, we use Israel as a case study, as its grid relies heavily on gas-fired power units, creating a strong interdependency between the electric system and the gas system.  The Israeli gas system currently operates with only two injection points and does not rely on imports or storage, at least at present. Any disruption or fault in the gas system necessitates nearly instantaneous actions on the electric system side to ensure an adequate supply of electricity during uncertain, pre-emergency, or emergency periods when gas becomes unavailable. Within this critical time-frame, operators must swiftly devise strategies to transition from primary fuel sources, typically gas, to secondary fuel sources, typically diesel.

The main contribution of this article is the introduction of a modular scheme model for emergency planning, in response to an unforeseeable gas shortage, that relies on a Markov Decision Process (MDP) to represent the inherent stochastic nature of the process.  This model is then used to address two fundamental questions. Firstly, what strategy should a power system operator adopt when operating dual-fuel generators, which might not be fully reliable, to transition from their primary to secondary fuel? Secondly, what are the implications of such a decision in terms of reliability, economic impact, and customer satisfaction?

This manuscript is structured as follows. In Section \ref{sec:formulation}, we elucidate how an optimal policy for a system-wide transition from primary to secondary fuel can be phrased as a MDP problem. Section \ref{sec:meth} details ways to tackle the solution of this problem and why, given the inherent complexity of this formulation, we delve into both approximate and heuristic methods to solve the problem. Section \ref{sec:investigation} outlines our empirical methodology, showcasing it through several parameterized use cases.  We conclude our discussion and highlight future directions in Section \ref{sec:conclusion}.

\section{Problem Formulation}\label{sec:formulation}

\begin{figure}[h!]
\center
\includegraphics[width=0.8\columnwidth]{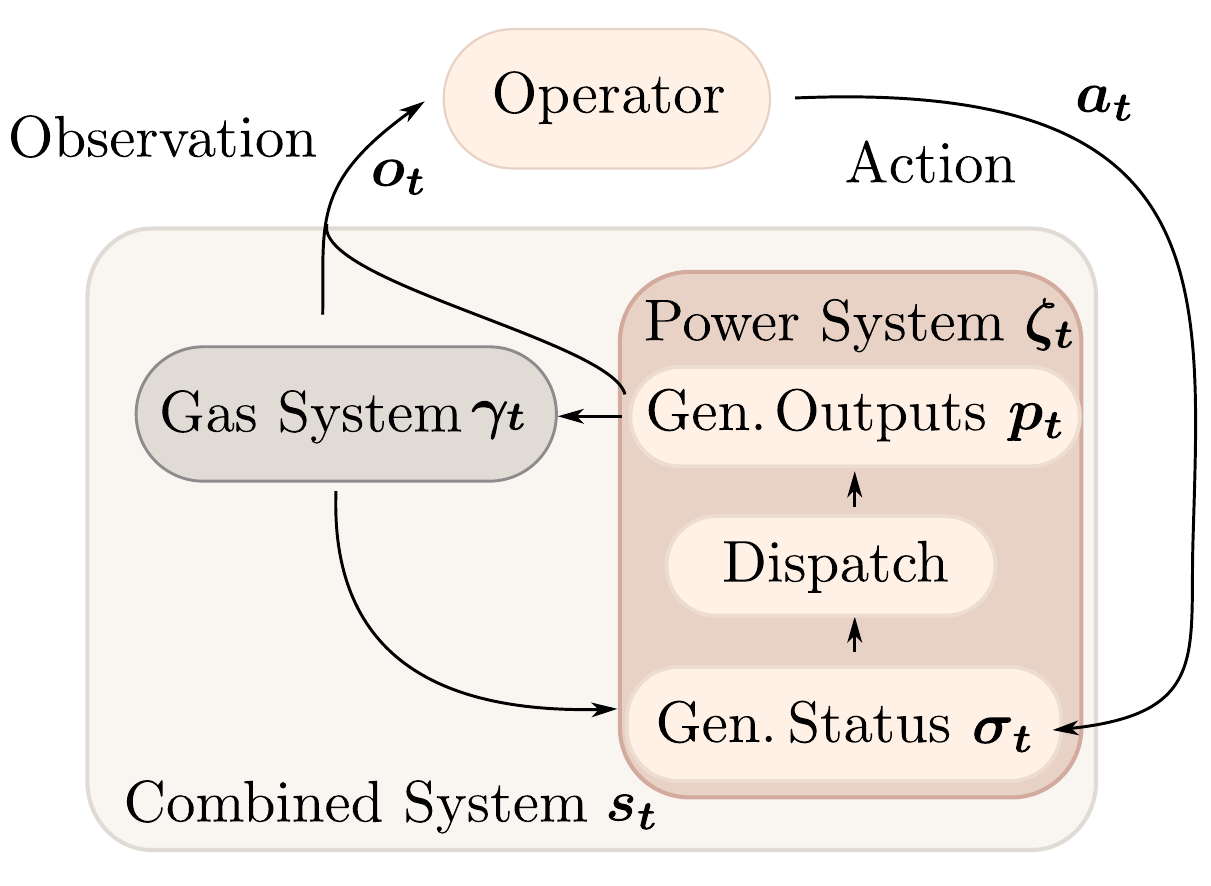}\caption{Diagram illustrating the relationships between system operator actions affecting generator status, the resulting states of the generators and gas systems, and the corresponding observations.}\label{fig:scheme}
\end{figure}

The situation we are studying in this work is an unforeseeable gas shortage. While, in theory, actions to mitigate its effects can be undertaken in both systems and jointly optimized, in many cases, the two systems are operated by two different companies which will want to minimize their own liabilities. Therefore, in the present work we focus on the actions that can be performed by a power system operator only. The schematic representation of the emergency plan  is shown in Fig.~\ref{fig:scheme}. Each block in Fig.~\ref{fig:scheme} can be modeled with more or fewer details. In the the following paragraphs, we detail our choices.

\paragraph{Operator's Actions \& Generators' Statuses}

\begin{figure}[t!]
\center
\includegraphics[width=1.0\columnwidth]{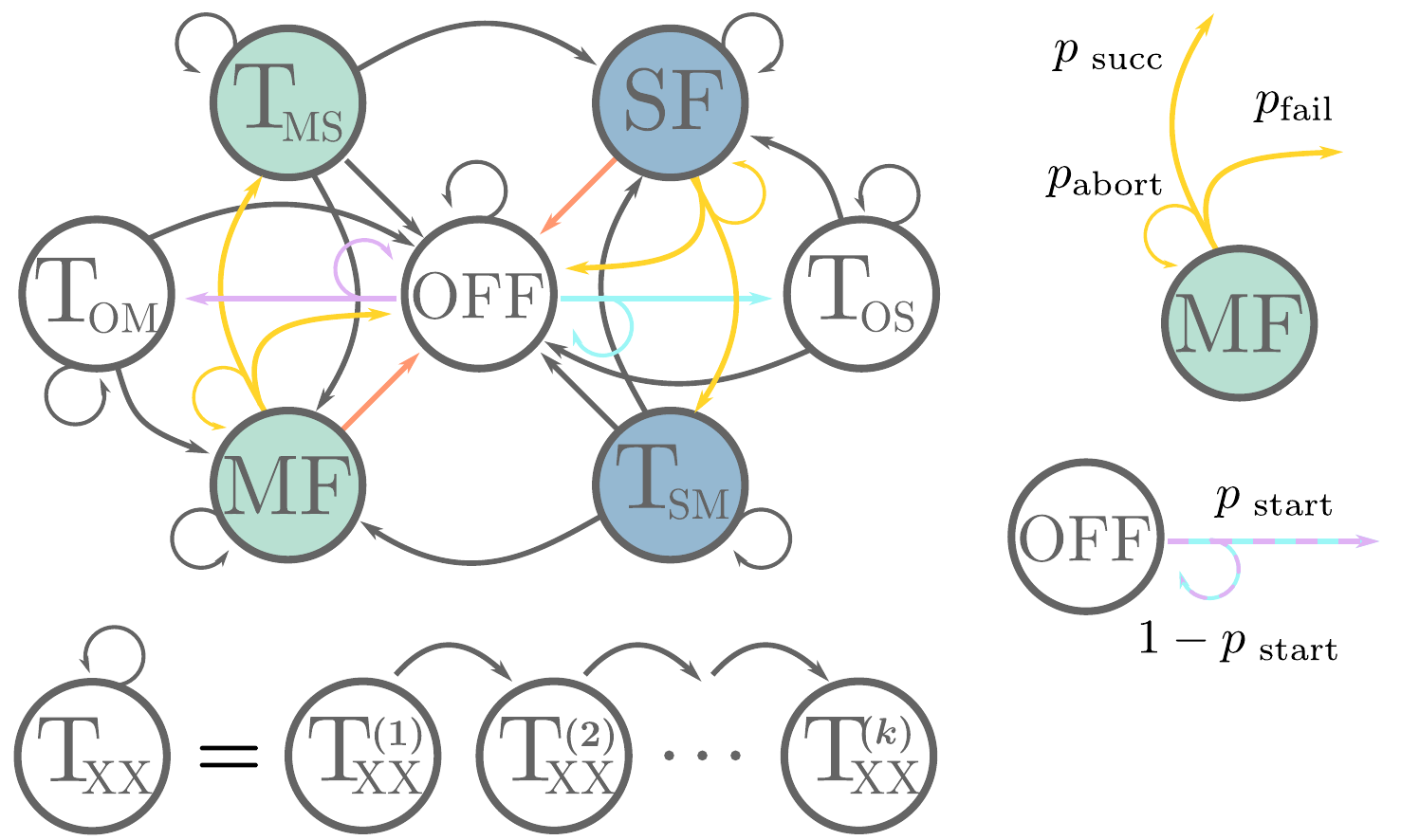}
\caption{
The Markov Process describes transitions between the operational states of a generator, which can be in one of three main states: Main Fuel (\textbf{MF}), Secondary Fuel (\textbf{SF}), and Offline (\textbf{OFF}). Additionally, there are transient states, e.g., (\textbf{OM}) and (\textbf{OS}), which represent transitions between \textbf{OFF} and \textbf{MF}, and between \textbf{OFF} and \textbf{SF}, respectively. These transient states consist of multiple sub-states (not shown), with the number of sub-states determined to match the expected transition duration based on the time step, $\Delta t$. The sub-script XX corresponds to the transient states (MS, SM, OS, or OM). States are color-coded based on their energy production status: green and blue states consume gas or diesel respectively. Actions are similarly color-coded: grey indicates no action, yellow represents a transition, purple and cyan denote start-up using main and secondary fuels respectively, and red represents shutdown. Shutdowns are considered instantaneous and reliable, with no transition state associated with them.
}\label{fig:markov}
\end{figure}

Our analysis simplifies the intricate dynamics of natural gas and electric power networks, focusing specifically on the management of dual-fuel units. The grid operator has the authority to instruct generators to switch between their primary and secondary fuels, initiate a shutdown, or start up. However, generators may fail to execute these commands. For instance, combustion instability could disrupt a smooth fuel transition and result in an unintended shutdown. To account for these uncertainties, we model the system dynamics as a Markov Decision Process (MDP), depicted in Fig.~\ref{fig:markov}. 

We consider the electric system's state $\bm{\zeta}_t$, which includes, among other factors, the statuses of generators, $\bm{\sigma}_t$. Each unit can be in one of three operational states, with four possible transition states between them. Detailed descriptions of these transitions are provided in the caption of Fig.~\ref{fig:markov}.

\paragraph{Generation Dispatch} 
The three blocks under the power system in Fig.~\ref{fig:scheme}, grouped together, can be interpreted as associated with a unit commitment task. This task is divided into two stages. First, the status of each generator is determined based on the operator's commands and the current state of the gas system. Next, the participating generators are dispatched optimally by solving a DC-Optimal Power Flow (DC-OPF) problem:
\begin{subequations}
\begin{align}
\min_{p_g,\, \xi_i}& \sum_{g} c_g(\sigma_g, p_g) + \sum_i c_i(\xi_i)\label{eq:opf_cost}\,,\\
&{\rm s.t.}\nonumber\\ 
& \forall g: p_g^{\min}(\sigma_g) \leq p_g \leq p_g^{\max}(\sigma_g) \,,\label{eq:opf_plim}\\
&\forall i: \sum_{j\sim i}b_{ij}(\theta_i-\theta_j) = \sum_{g \in G_i}p_g - d_i + \xi_i\,,\label{eq:pf1}\\
&\forall j\sim i: -f_{ij}^{\max} \le b_{ij}(\theta_i-\theta_j)\le f_{ij}^{\max}\,,\label{eq:pf_lim}
\end{align}
\label{eq:opf_prob}
\end{subequations}
where $p_g$ represents the dispatch of generator $g$, $p_g^{\rm min}$ and $p_g^{\rm max}$ are its limits, and $c_g$ is the associated cost function. Here, $b_{ij}$ is the susceptance of the line between buses $i$ and $j$, $f_{ij}^{\rm max}$ is the thermal limit of the line, $d_i$ is the load at bus $i$, and $\theta_i$ is its phase angle. The inclusion of nodal load shedding terms $\xi_i$ and their associated costs $c_i$ ensures that the optimization problem \eqref{eq:opf_prob} remains solvable. Temporal dependencies are not explicitly shown in Eq.~\eqref{eq:opf_prob}, with subscripts $t$ and summations over time omitted for clarity. The dependency on $\sigma_{g}$ in Eqs.~\eqref{eq:opf_cost} and \eqref{eq:opf_plim} reflects the fact that generator costs vary based on whether they are running on primary fuel (natural gas) or secondary fuel (diesel).

\paragraph{Gas System} Once actions have been implemented, and generators' outputs set, the gas system is evolved for the duration of a time step $\Delta t$. Assuming that the states of the gas and power systems at time $t$ are labeled $\bm \gamma_t$ and $\bm \zeta_t$ respectively, the state of the gas system is updated as $\bm \gamma_{t+1} =\bm F(\bm \gamma_t,\, \bm \zeta_t)$. If the function $\bm F$ is a proper description of the system dynamics, it asks for the solution of a system of PDEs and hence the use of a dedicated simulator~\cite{hyett2023control}. As described in \cite{CEC2019}, the generators' fuel intakes $f_g$ is obtained from their outputs $p_g$, by the heat rate curve

\begin{equation} \label{eq:heat-rate-curve}
f_g = \alpha_0 + \alpha_1 p_g \big/p_g^{\max} + \alpha_2 \Big(p_g \big / p_g^{\max}\Big)^2,
\end{equation}
where $\alpha_0, \alpha_1$, and $\alpha_2$ are the coefficients of a fitted quadratic function, which may be uniquely assigned for each unit in the system. Subsequently, fuel intakes $f_g$ are consolidated by station. As shown in Fig.~\ref{fig:scheme}, generator states $\bm \sigma_t$ can be forcibly changed if $\bm \gamma_t$ satisfies certain conditions, e.g. if the pressure at a substation falls below a threshold, $\pi_s < \pi_s^{\rm min}$, the generators supplied by this substation are forced to be shut down instantaneously. 

In the spirit of the copperplate model, introduced below, the gas system dynamics can also be approximated by only focusing on linepack changes. In this case, $\bm F$ reads
\begin{equation}
l_{t+1} = l_t - \Delta({\bm\zeta}_t)\,,
\end{equation}
where $\Delta({\bm\zeta}_t)$ is the amount of gas consumed during the time step. Here we define linepack in GWh; this conversion assumes a constant of proportionality between mass of gas consumed and energy generated. An efficiency factor $\eta_g = 0.40$ is used, leading to $ \Delta({\bm\zeta}_t)=\sum_{\sigma_{g;t}\in\{\textbf{MF}, \textbf{T}_\textbf{MS}\}}\eta_g^{-1} p_{g;t}\Delta t$. The linepack's depletion switches units in \textbf{MF} or $\textbf{T}_\textbf{MS}$ to \textbf{OFF}.

\subsection{Israeli Infrastructure}

\paragraph{Copperplate System} We assume that the Israeli power system sees no congestion. Consequently, constraint \eqref{eq:pf_lim} can be neglected. This has the subsequent effect that loads $d_i$ and load shedding terms $\xi_i$, defined in problem~\eqref{eq:opf_prob}, can be gathered into global quantities, $d_t$ and $\xi_t$ respectively.

\paragraph{Simplified Dispatch}
The fact that roughly 40\% of the country's generation capacity is still state-owned allows us to simplify the generation dispatch and bypass the resolution of an OPF. The following simplified dispatch scheme is used
\begin{equation}\label{eq:pg}
p_{g;t} = \varepsilon_t\, p_g^{\rm max}  + (1 - \varepsilon_t)\,p_g^{\rm min}\,,\; \forall g \in \mathcal{G}\,,
\end{equation}
with $\varepsilon_t$ is a global signal varying between 0 and 1, balancing between maximum and minimum power output levels. Its value is set so that generation meets demand, if possible. 

\paragraph{Homogeneous Fleet} For the sake of simplicity and interpretability, we assume that all generators are dual-fuel units; other generators were factored out. We further assume that they all belong to the same  reliability class, defined by their transition probabilities dictating the behavior of the MDP, which are displayed in Table \ref{tab:prob} .
\begin{table}[h!]
    \centering
    \begin{tabular}{lcccc}
    \hline
    &$p_{\rm abort}$ & $p_{\rm succ}$ & $p_{\rm fail}$ & $p_{\rm start}$\vphantom{$f_f^f$}\\
    \hline
    super-reliable & 0.01 & 0.98 & 0.01 & 0.98\vphantom{$f_f^f$}\\
    reliable & 0.05 & 0.90 & 0.05 & 0.90\\
    fairly reliable & 0.10 & 0.80 & 0.10 & 0.80\\
    unreliable & 0.15 & 0.70 & 0.15 & 0.70\vphantom{$f_f^f$}\\
    \hline
    \end{tabular}\vspace{0.1cm}
    \caption{Transition probabilities for different generator classes.}\label{tab:prob}
\end{table}

\begin{figure}[h!]
\center
\includegraphics[width=0.8\columnwidth]{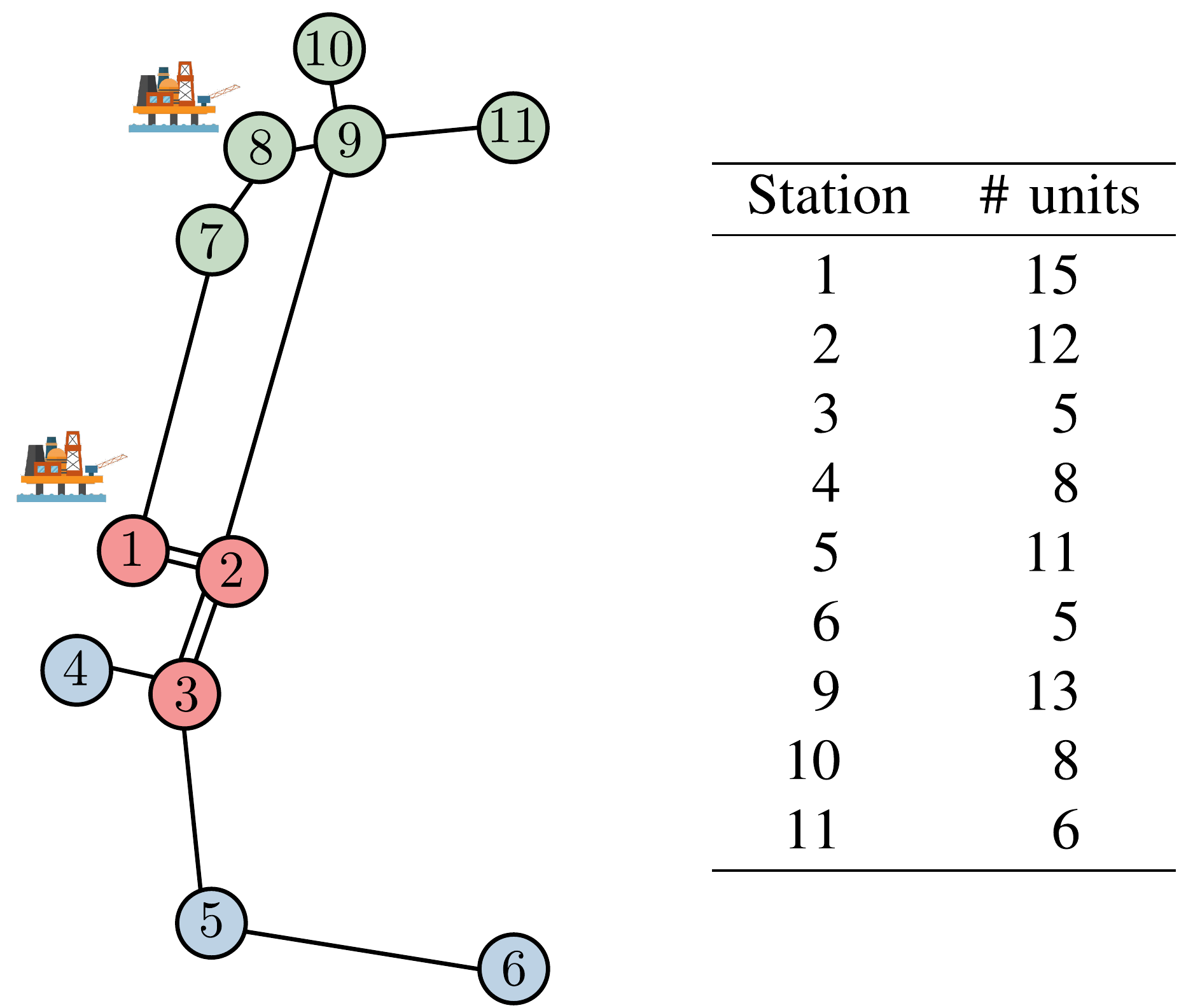}
\caption{Map of the Israeli gas system and list of units. Gas platforms denote the two injection points, which are lost at the start of the emergency scenario. The three regions used in Fig.~\ref{fig:comparisons}: North (green), Center (red) and South (blue).}\label{fig:map}
\end{figure}

\paragraph{System Overview} We use a synthetic model of the Israeli infrastructure, consisting of 83 dual-fuel 150MW which are supplied from one of 11 nodes of a simplified natural gas transmission system, see Fig.~\ref{fig:map}. Generation and load shedding costs are assumed to be linear. Unserved energy is priced at \$20,000/MWh.
Marginal costs of primary and secondary fuels are set at \$30/MWh and \$420/MWh. These prices are consistent with those present in Israel's real energy market. We assume that it takes 20 minutes for all generators to transition between fuel types, or to start up. We have set the pressure $\pi_s^{\rm min}$ at which they can withdraw gas from the substation at 50bar. 

\paragraph{Complexity of Actions} Transitioning of generators is limited by the number of available personnel, prohibiting simultaneous transitions of multiple units. To mitigate this limitation, we set a maximum number of allowable actions, $\#\bm a_t\le K$, $K\in\{1,2,3,\ldots\}$, for the system operator at any point in time.

Finally, the objective function can be concisely phrased as
\begin{subequations}
\label{eq:opt}
\begin{align}
\min_{{\bm a}_1,\ldots, {\bm a}_{T}} & \mathbb{E}\bigg[\sum_{t=1}^{T}\Big( \xi({\bm \sigma}_{t},{\bm \gamma}_t) +\sum_{g\in\mathcal{G}}c(\sigma_{g;t};p_{g;t})\Big) \bigg],\\
&\text{s.t.}\nonumber\\
\bm \sigma_{t+1} &= \bm T(\bm a_t, \bm \sigma_t,  \bm \gamma_t)\,,\\
\bm \gamma_{t+1} &= \bm F(\bm \gamma_t,\, \bm \zeta_t)\,,
\end{align}
\end{subequations}
where the function $\bm T$ gathers the random drawing of the new states $\bm \sigma_t$ based on the MDP and the projection due to conditions met by $\bm \gamma_t$.

Before moving on, we would like to emphasize that the assumptions and simplifications made here are, by no means, shortcomings of the model we present in this work. They only allow us to develop and to rapidly validate our methodology, which can seamlessly be extended to richer, more complex systems. However, the interpretation of the results as general guidance to system operators, which is the intention of this preliminary work, would be more difficult.

\section{Methodology}\label{sec:meth}

Our approach streamlines a complex stochastic optimization challenge by integrating key factors into the objective function. These include: (a) limiting simultaneous transitions of dual fuel generators, (b) optimizing gas re-distribution along the system (subject to pressure limits), (c) minimizing operational costs, (d) reducing switching efforts, (e) delaying the switch to secondary fuel to hedge against premature gas supply restoration, and (f) curtailing the load shedding costs. This encapsulates a risk-economy trade-off and addresses the inherent stochastic nature of fuel transitions.

To illustrate the inherent uncertainties and challenges in the decision-making process, let us examine a simplified example. Suppose the operator perceives an immediate requirement for a swift transition to secondary fuel. An initial, straightforward response might be to instruct all dual-fuel generators to switch fuels simultaneously and promptly. Should this transition occur seamlessly and efficiently, it would be considered an ideal outcome. Nevertheless, this rapid and uniform approach carries significant risks. A failure in one or more units during such an abrupt transition could lead to a decrease in the power supply, causing insufficient energy distribution to consumers, and may necessitate load shedding, which is highly undesirable.

On the other hand, if the operator opts for a cautious approach, transitioning one unit at a time, this method might still lead to load shedding. In a simplified scenario, the linepack -- the gas volume within the system -- might be depleted before completing the transitions. In a more complex system analysis, which takes into account the inherent dynamics of the natural gas system, generators might need to be disconnected due to pressure drops below acceptable limits at their associated gas stations.

Therefore, the ideal strategy likely falls between these extremes, necessitating a balanced approach that carefully navigates the trade-offs. This project aims to understand these complexities and devise a decision-making framework that offers optimal and practically viable solutions to the operator.

\paragraph{Dynamic Programming Approach} 
Dynamic Programming (DP) offers a theoretically robust framework to solve MDPs given known transition probabilities. However, DP faces computational challenges, often becoming prohibitively expensive as the dimensionality of state and action spaces increases -- a situation typical in power systems with numerous generators. Particularly demanding is the incorporation of constraints on transition numbers per step, such as $K \leq 10$. To respect this limitation, we identify functions of primary variables from exponentially large sets, emphasizing the need for practical, albeit approximate, solutions as presented in this study.

\paragraph{Parameterized Markov Processes} 
For this research, we generate a sequence of actions to be followed by the system operator from a parameterized Markov Process, a pragmatic approach that generates and evaluates multiple viable action plans based on rule-based selections. Although this method may not guarantee the discovery of an optimal solution, it facilitates the iterative refinement of strategies and inspires the development of novel alternatives. While Parameterized Markov Processes (PMP) may not achieve an optimal solution, it is still successful at mitigating load shedding in the event of an emergency, when convergence to optimality may not be sought. Moreover, the parameters, and thus the ``correct'' heuristics, may be learned from experience rather than fixed. Such an evolving policy may converge to optimality if recovered through Reinforcement Learning (RL) \cite{sutton2018reinforcement}. An in-depth analysis of PMP and its implications will be further discussed. In Section \ref{sec:line-pack}, we adopt a simplified approach to model the gas system, focusing solely on its linepack (overall current capacity). This abstraction facilitates the exploration of a broader range of scenarios and hyper-parameter settings, albeit in a less detailed manner. Subsequently, in Section \ref{sec:network}, we narrow our focus to the most realistic scenarios and employ a detailed model to simulate gas flow across the network. This involves solving the Partial Differential Equations (PDEs) that characterize gas flows for each generated sample path in the Markov Process (MP).
\begin{algorithm}\label{alg:MP}
\SetKwInOut{Input}{Input}
\caption{Prescribed emergency plan at time $t$.}
\Input{Maximal number of actions $K$ and reserve capacity $R$;}
$\textit{action} \gets 0$\;
\While{action $ < K$}{
 \eIf{available capacity $<$ demand + reserve}{
   start up an offline unit on secondary fuel;
 }{
   transition a main fuel unit to secondary;
 }
 $\textit{action} \gets \textit{action}+1$\;
}
\end{algorithm}

\section{Numerical Investigation}\label{sec:investigation}

As outlined earlier, we convert the MDP Eq.~\eqref{eq:opt} into parameterized Markov Process to explore different operational scenarios. This approach allows us to examine a subset of the decision-making elements of the original MDP. A time resolution of $\Delta t = 5$min was chosen for these experiments.

\subsection{Linepack Limited Emergency Policy}\label{sec:line-pack}

In this first set of simulations, the two systems are approximated by ``copperplate'' models. In that setup, the two meaningful control parameters are the maximum number of actions $K$ per time step the operator can perform and the reserve capacity $R$ which acts as a fail-safe during transition failures.  In order to study their effect, the fundamental MP rules are detailed in Alg.~\ref{alg:MP}. The available capacity is defined by adding up the capacity of available units as $ \sum_{\sigma_{g;t}\in \mathcal{G}}p_g^{\max}$. We simulate $10,000$ runs for each parameter set. In all the following figures with shaded areas, the solid line represents the mean value and the areas display, for the darkest to the lightest, 90\%, 98\% 99.8\% and the complete observed range. The initial linepack $l_0$, indicating the primary fuel available at the emergency's start, is assumed to be at 60GWh unless otherwise stated. 

\begin{figure}
\includegraphics[width=\columnwidth]{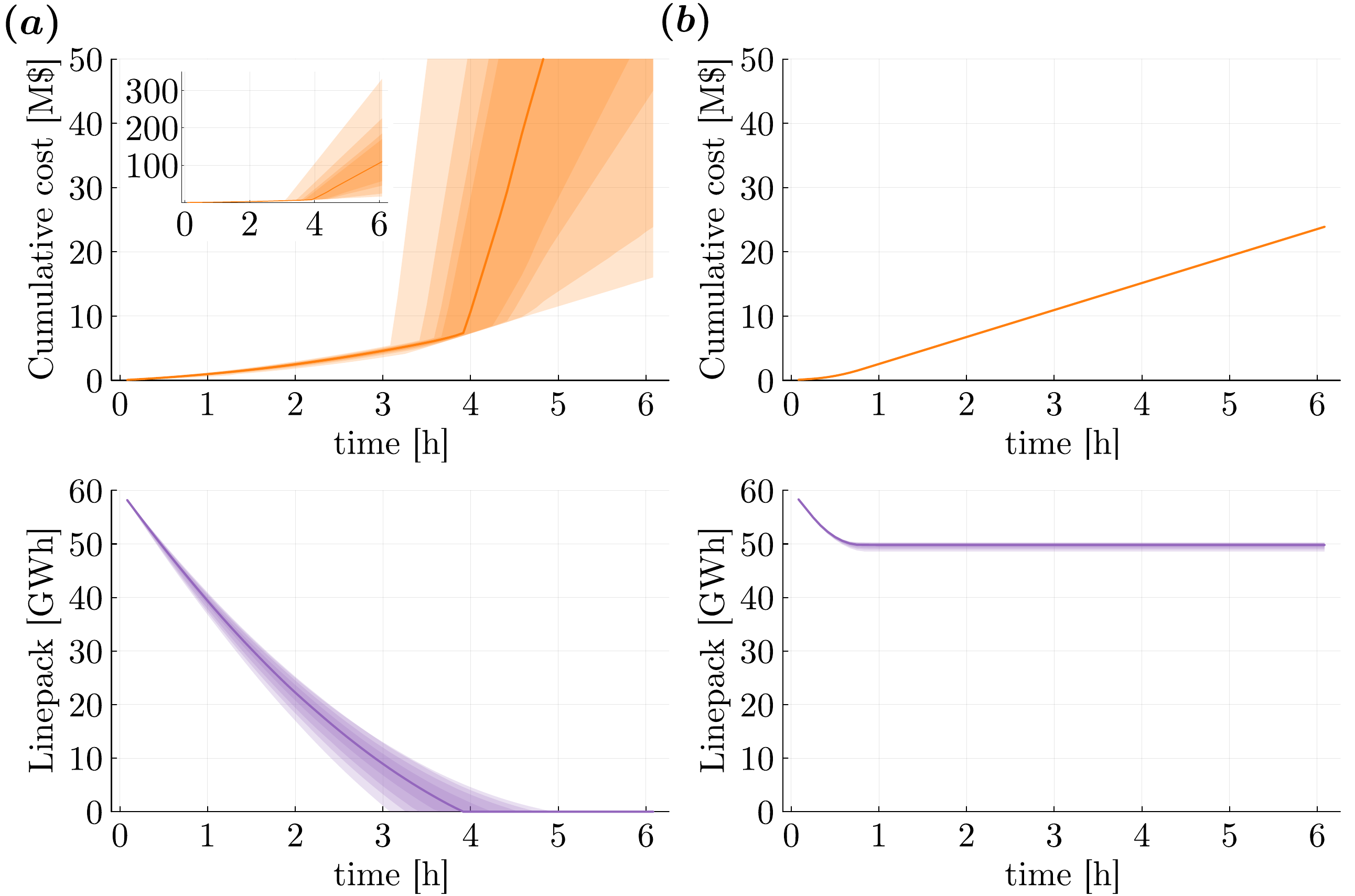}
\caption{Scenario with super reliable units and $K$ is set to 1 (column a) and to 10 (column b) and the reserve is $R=500$MW.} \label{fig:super_10}
\end{figure}

The first factor we examine is the impact of limiting the maximum number of actions available at each time step. Although we do not explicitly quantify the complexity of emergency plans, simpler policies are generally preferred. The left column of Fig.~\ref{fig:super_10} presents the results for super-reliable units when only one action is available. In this scenario, the linepack is depleted, leading to a rapid increase in system costs. Conversely, as shown in the right column, allowing the operator to perform 10 actions results in a faster transition and preserves most of the linepack. 

Since secondary fuel is more expensive than the main fuel, it would be expected that option (b), with $K=10$, leads to a higher cumulative cost compared to option (a), where only one action is allowed, as more units are likely to switch to secondary fuel. However, there is a small probability, approximately 1 in 10,000, that the transition completes before gas depletion in option (a), making it potentially cheaper in such rare cases. This motivates the system operator to consider protocols with a high $K$, but still impose an upper limit to prevent unnecessary transitions to secondary fuel.

\begin{figure}[h!]
\includegraphics[width=\columnwidth]{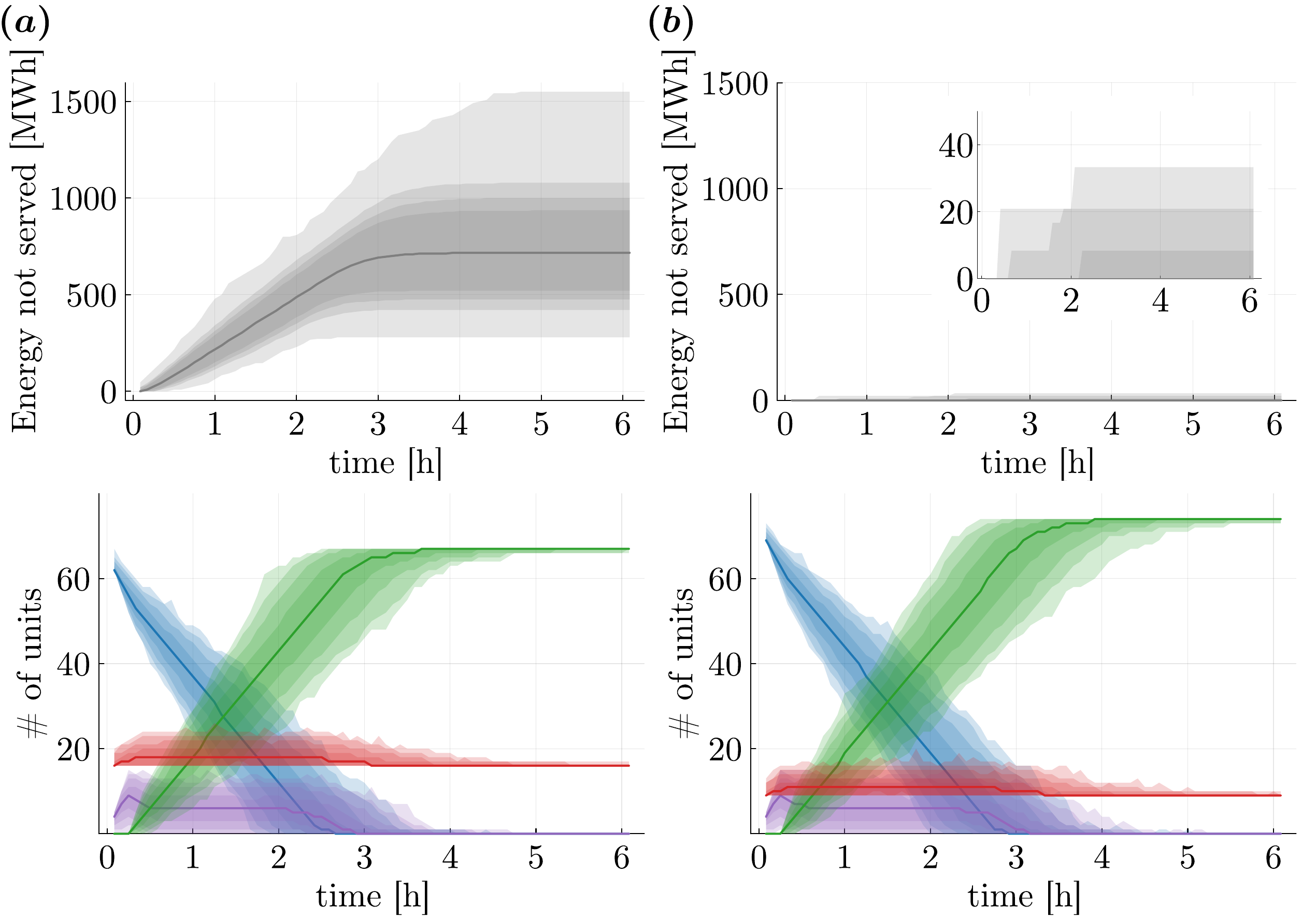}
\caption{Scenario with fairly reliable units and max number of actions $K$ is set to $5$ and with no reserve capacity $R=0$MW in (a) and with a reserve capacity $R=1000$MW in (b). The bottom plot shows the status of the fleet: the number of generators  on main fuel, on secondary fuel, on transition, and offline are displayed in blue, green, purple and red, respectively.}\label{fig:fairly_4}
\end{figure}

Next, we analyze the impact of varying reserve capacity $R$ for fairly reliable units. In column (a) of Fig.~\ref{fig:fairly_4}, no reserve is allocated. As expected from the startup and transition success probabilities, the fairly reliable units experience more frequent failures compared to the super-reliable units used in Fig.~\ref{fig:super_10}. Consequently, the overall state of the fleet becomes more uncertain; for example, in some cases, the number of offline units exceeds the average by at least five units, which inevitably leads to load shedding. Naturally, maintaining a reserve comes at a cost, as more units must participate in the dispatch. 

In column (b), the reserve capacity is set to 1 GW, allowing approximately six units to fail before load shedding becomes necessary. Notice that the number of generating units is higher compared to column (a). Interestingly, apart from a slight increase in the number of units using secondary fuel, the transition unfolds almost identically in both cases. However, when considering the energy not served -- defined as the integral of load shedding over time -- the two scenarios differ significantly. In option (b), load shedding is avoided in nearly all simulations, and when it does occur, its magnitude is minimal.

\begin{figure*}[h!]
\includegraphics[width=\textwidth]{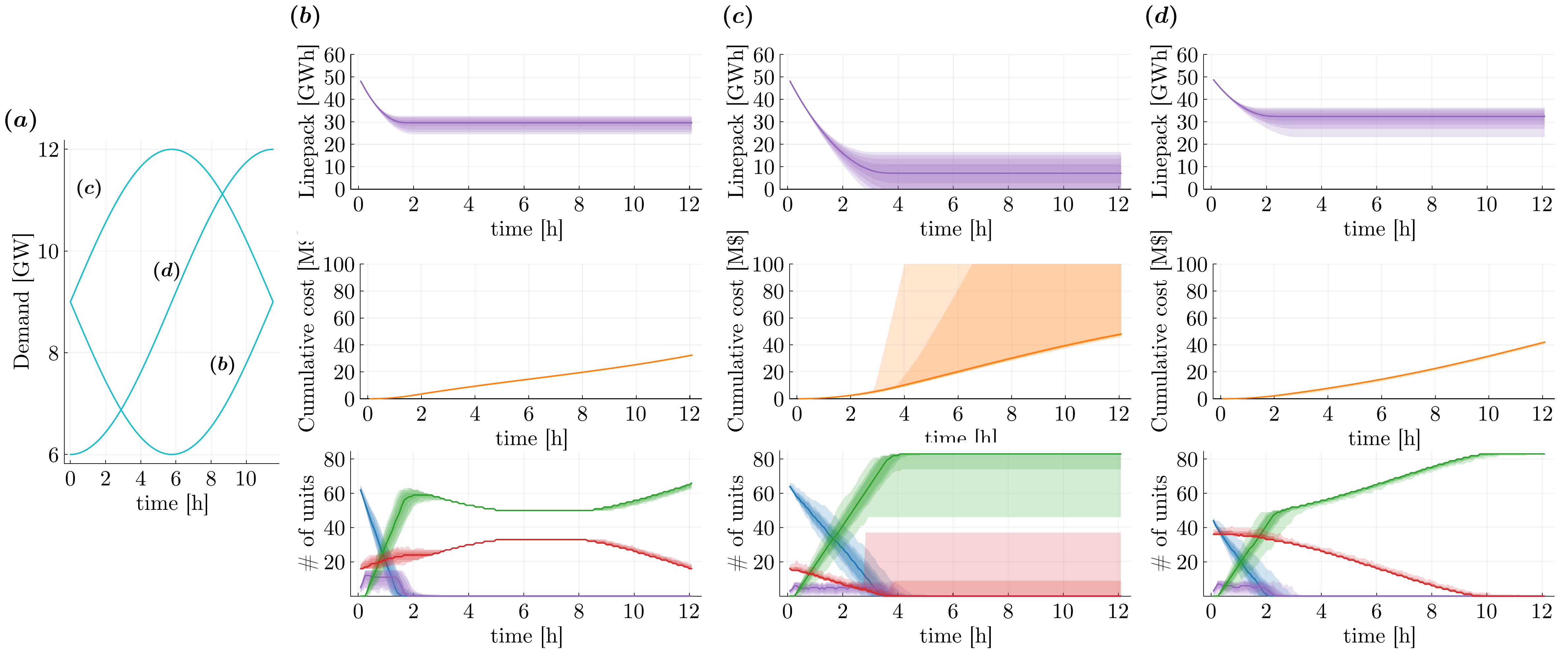}
\caption{(a) Three generic load patterns: \textit{Nighttime}, \textit{Noon} and \textit{Morning}. (b)--(d) the corresponding results for reliable units with $K=3$ available actions and a reserve $R=1$GW.}\label{fig:varying}
\end{figure*}

Our first two use cases illustrate the impact of the number of available actions and reserve levels, as well as the effect of unit reliability class under a fixed demand profile. We now shift focus to examine the influence of varying demand, as depicted in Fig.~\ref{fig:varying}. To explore this, we consider three synthetic demand curves: \textit{Nighttime} -- initially decreasing, then rising, \textit{Noon} -- initially increasing, then decreasing, and \textit{Morning} -- characterized by consistently rising demand. At first glance, \textit{Morning} appears to be the least opportune time for a gas shortage, given the rapidly increasing demand. However, since the system demand is low at the onset of the shortage, a significant portion of the fleet is offline and can be started directly on secondary fuel. 

In contrast, \textit{Noon} proves to be the most challenging scenario. Higher initial demand means that more units are running on main fuel and need to transition to secondary fuel. In some instances, the linepack depletes, leading to load shedding. Under such circumstances, it is advantageous to allow a larger number of simultaneous actions, e.g., $K \geq 5$. \textit{Nighttime}, though it also begins with most units on main fuel, is relatively uneventful as the demand decreases. By the time demand rises again, the transition process is already underway. It is worth noting that the precise course of the transition depends not only on the demand profile but also on the state of the gas system and control parameters in the PMP.

\begin{figure}[h!]
     \centering
     \includegraphics[width=0.95\columnwidth]{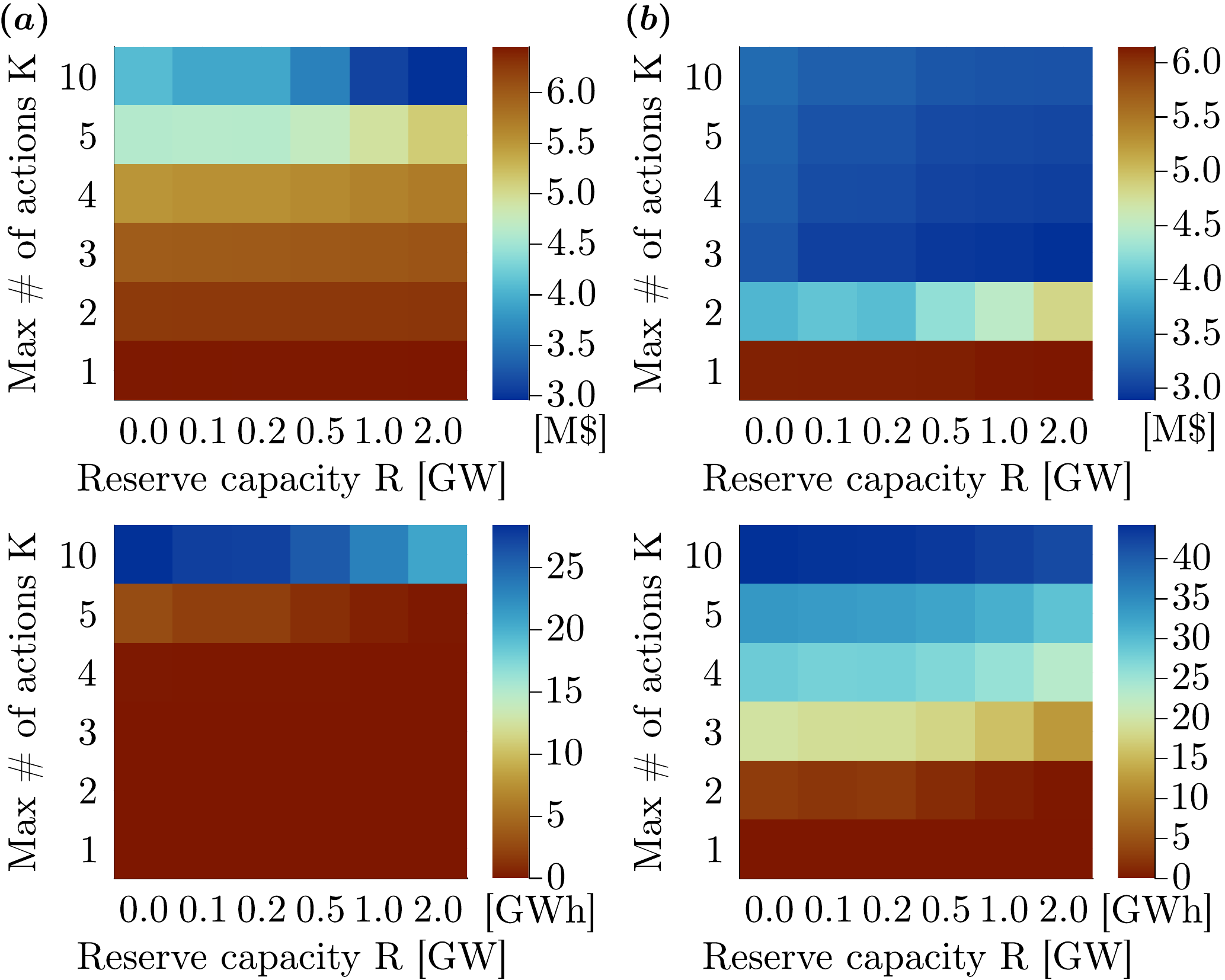}
     \caption{Columns (a) and (b) show the average cumulative costs in millions of dollars (top)  and remaining linepacks in GWh (bottom) for unreliable and reliable units respectively.}\label{fig:unreliable}
 \end{figure}

Thus far, we have changed at most one control parameter per use case. As a last study for this copperplate model, we search for the optimal values of $R$ and $K$.
With Fig.~\ref{fig:varying}, we have seen their optimal values depend on the load profile under consideration. In Fig.~\ref{fig:unreliable}, column (a) shows the level of the linepack at termination, and the total cost over the episode for unreliable units. We observe that, for this class of units, it is almost impossible to prevent the linepack from being depleted. With respect to cost, the cheapest strategy is to maximize both the number of allowed actions and the reserve capacity. On the other hand, if the units are more reliable, the cheapest strategy occurs at lower reserve and number of actions. These findings are aligned with those in Figs.~\ref{fig:super_10} and \ref{fig:fairly_4} and their respective commentary. If the operator's objective is a trade-off between minimizing the costs and preserving linepack, the optimal strategy is inclined towards allowing more actions.

It is worth noting that, for the unreliable units in the first row of Fig.~\ref{fig:unreliable}, there are some cases for $K \leq 5$ in which the cumulative cost increases as the reserve increases. It is intuitive to expect that an increase in reserve capacity should protect against load shedding, thereby reducing the cumulative cost. Thus, the increased cost likely stems from a greater number of units running on secondary fuel. This is consistent with the linepack observations in the second row, as running units on secondary fuel reduces the need for main fuel (natural gas) to be depleted. Nonetheless, in all cases and for all reliability, the lowest costs are observed when the most actions are allowed, e.g., $K = 5,10$. Indeed, for $K=10$, the expected decrease in cost occurs as reserve increases. This substantiates the point made in the previous paragraph, i.e., that the optimal policy involves maximizing the number of allowed actions for the system operator.

\begin{figure*}[t!]
\centering
\includegraphics[width=1.0\textwidth]{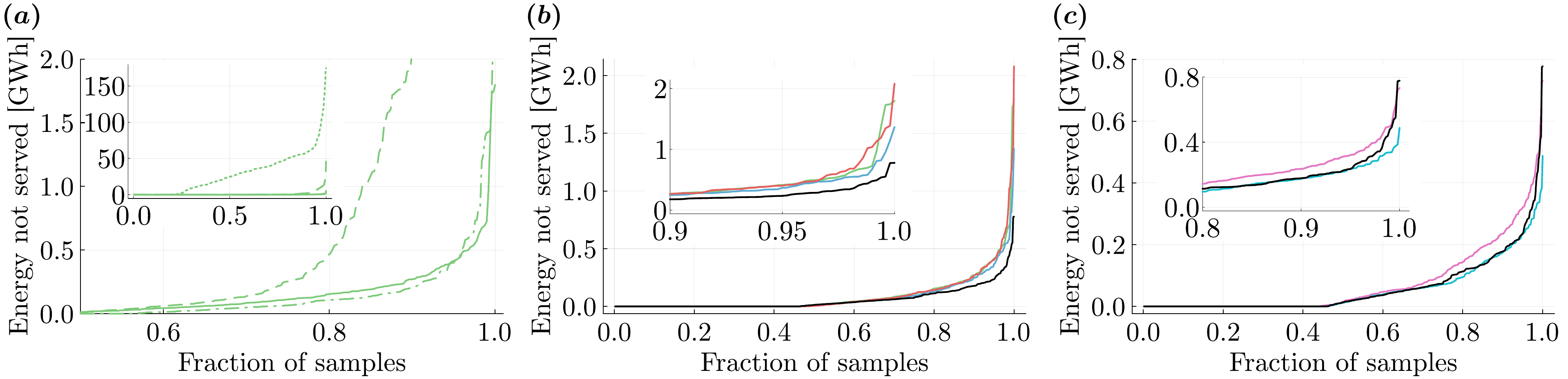}
\caption{Comparison of the average amount of energy not served, in GWh for various strategies. The horizontal axes provide the percentage of observations below the corresponding values on the vertical axes. In panel (a), a strategy favoring transitioning units in the North is employed, and the number of allowed actions $K$ is varied: 2 (dotted), 3 (dashed), 4 (dashdotted) and 5 (solid). In panel (b), strategies favoring the North (green), Center (red) and South (blue) are compared against a random selection strategy (black). In panel (c), strategies prioritizing transitions for sites with low (cyan) and high (pink) pressure are compared against the same random selection strategy. The number of allowed actions in panels (b) and (c) is set to 5.
\label{fig:comparisons}}
\end{figure*}

\subsection{Gas-Network-Aware Emergency Policy}\label{sec:network}

The approach described in Section \ref{sec:line-pack} simplifies the assessment by globally evaluating linepack, neglecting the specific gas pressures at individual nodes. To more accurately reflect the dynamics between the power and gas networks, we integrate the real-time resolution of gas flow equations into our decision-making framework.

The interaction between the power and gas systems is monitored by concurrently logging demand and tracking the gas system's state evolution. Additionally, various scenarios reflecting different transition strategies are examined, influenced by the geographic distribution and reliability of generation units throughout the network:
\begin{description}
   \item[1A] do not transition any units;
   \item[1B] randomly select units for transition;
   \item[2A] select units in the north first, then randomly;
   \item[2B] select units in the south first, then randomly;
   \item[2C] select units in the center first, then randomly;\
   \item[3A] select units supplied by gas stations with the lowest pressures first;
   \item[3B] select units supplied by gas stations with the highest pressures first.
\end{description}
The strategies are chosen to highlight effective intuition-guided protocols, and to contrast them with their counterparts. For example, in the event of a pressure fallout, it is advantageous to retain units associated with higher pressure gas stations, as they can withstand the fallout for a longer duration than those units associated with lower pressure gas stations. This motivates strategy 3A, prioritizing lower pressure units for transitions, with strategy 3B included for comparison.

For each of these strategies, we apply the policies 500 times and record the cumulative amount of load shedding, in GWh, for each iteration. This quantity measures the relative effectiveness of each strategy in serving the demand of the Israel power system. The distributions of the load shed for the different scenarios are depicted in Fig.~\ref{fig:comparisons}. These simulations reveal the sensitivity of the underlying Markov Process to the selection criteria, as well as to the parameters governing the actions of the system operator. In particular:

\paragraph{Number of allowed actions}
In Fig.~\ref{fig:comparisons} (a), Strategy 2A is assessed when at most $K$ units are allowed to be transitioned at once, with $K = 2,3,4,5$. It is evident that as the number of allowed actions increases, the amount of load shedding decreases. While the increased flexibility in increasing $K$ is intuitive, a closer examination of the sensitivity to $K$ is required to determined the threshold value of $K$, after which load shedding is reduced to nearly zero.

Determining this threshold is critical for system operators to implement the suggested strategies successfully, as significant load shedding can still occur under seemingly moderate restrictions on $K$. The results suggest that the system operator should be permitted to simultaneously transition $K \geq 5$ units in order to achieve adequate reduction of load shedding. Finally, a similar trend with the increase of $K$ is observed for all of the other suggested policies, so the guidance is widely applicable regardless of whether a random policy or a rule of thumb is chosen by the system operator.

\paragraph{Geographic preference}
In Fig.~\ref{fig:comparisons} (b), the geographically motivated strategies of prioritizing units in the North, South, and Central regions of Israel for transitions are compared as Strategies 2A, 2B, and 2C, respectively. The geographic structure of the power system is not itself sufficient to infer which region should be favored to result in a stronger policy. However, prioritizing transitions in the South is expected to be more favorable, as minimum pressure constraints are more likely to be challenged in this regime. Indeed, Strategy 2B, which favors transitioning units in the South, achieves a higher probability of (nearly) avoiding load shedding than favoring the north in Strategy 2A or center in strategy 2C. However, the upper tail of the load shedding distribution from 2B displays larger values than for the random policy, demonstrating that the strategy is more vulnerable to high degrees of load shedding for selected scenarios. Furthermore, the performance of the three regional strategies are comparable for a majority of the samples, with the policies most distinguishable at their upper tails. To better effectuate Strategy 2B, we may introduce additional reserve to the system to mitigate the outliers observed in the present data.

\paragraph{Gas pressure preference}
In Fig.~\ref{fig:comparisons} (c), we evaluate the gas pressure levels at stations containing the generators. In Strategies 3A and 3B, priority is given to transitioning generators associated with low and high-pressure stations, respectively. Intuitively, transitioning low-pressure units seems more advantageous, as these units are at greater risk of forced shutdown if their station's pressure falls below the minimum threshold.
The results confirm that Strategy 3A, which favors transitioning low-pressure generators, generally performs better. This is particularly evident in the upper quarter of the examined distributions. However, the two strategies show similar performance across a larger proportion of samples than expected. This observation suggests that pressure-based heuristics may not be as straightforward to identify, which strengthens the argument for random generator selection, as discussed in the following paragraph.

\paragraph{Effectiveness of random policy}
In panels (b) and (c), Strategy 1, corresponding to a random selection criterion for generator transitions, with all units in play, is shown in black for comparison. It should be noted that the random policy is overall effective, and is indeed more effective than any of the regional policies 2A-2C, achieving the greatest probability of near-zero load shedding. The success of the random policy is not surprising, as the model of the Israel system assumes that all generators are homogeneous; i.e., their upper and lower dispatch limits are the same, and the coefficients of the heat-rate curve in Eq.~\eqref{eq:heat-rate-curve} are also identical.  
Moreover, the system is assumed to be ``copperplate'' without transmission constraints, further increasing the homogeneity of selecting units for transition. Under these assumptions, the relative success of the random policy justifies the use of a probabilistic decision-making procedure in unit commitment. However, as the model is enriched and realistic parameters and attributes are learned, rules of thumb in the spirit of Strategies 2A, 2B, 2C, 3A, and 3B may be developed that possess an inherent advantage to a random policy. This is already evident in Strategy 3A, whose relative success may be interpreted as a consequence of the primary distinguishing factor between the power units, the pressures of their supplying gas stations. This further emphasizes that the gas dynamics must be faithfully accounted for in the decision-making process on the power system side.

\section{Conclusion and Path Forward}\label{sec:conclusion}

This project seeks to equip power-system control room operators with tools tailored to aid in formulating strategies to address critical or near-emergency scenarios demanding swift decision-making. While our goal is not to replace operators, we strive to provide them with computational tools that streamline decision processes in situations where natural gas resources are scarce and unreliable, thereby increasing the risk of load shedding. Our solutions prioritize simplicity, cost-effectiveness, trustworthiness, and explainability. To the best of our knowledge, this manuscript is the first to frame the problem, formulate it using a Markov Decision Process (MDP), and then suggest an intuitive but reduced approach to its resolution by evaluating multiple parameterized Markov Processes (MPs). While we do not claim to provide a definitive finite solution, we pave the way for a holistic resolution to these challenges. Future work will involve generalizing to a fully constrained optimal power flow model as described in Section \ref{sec:formulation}. Other simplifying assumptions, such as the homogeneity and identical performance of generators, will also be relaxed. As well, efficient incorporation of the gas dynamics, potentially including compressors, will be completed. Finally, the heuristic-driven policies will evolve into a learned optimal policy via RL. In total, these adjustments will enable broad applicability and scalability of our method to a variety of interconnected power and gas systems.

\bibliographystyle{IEEEtran}

\begin{thebibliography}{10}
\providecommand{\url}[1]{#1}
\csname url@samestyle\endcsname
\providecommand{\newblock}{\relax}
\providecommand{\bibinfo}[2]{#2}
\providecommand{\BIBentrySTDinterwordspacing}{\spaceskip=0pt\relax}
\providecommand{\BIBentryALTinterwordstretchfactor}{4}
\providecommand{\BIBentryALTinterwordspacing}{\spaceskip=\fontdimen2\font plus
\BIBentryALTinterwordstretchfactor\fontdimen3\font minus
  \fontdimen4\font\relax}
\providecommand{\BIBforeignlanguage}[2]{{%
\expandafter\ifx\csname l@#1\endcsname\relax
\typeout{** WARNING: IEEEtran.bst: No hyphenation pattern has been}%
\typeout{** loaded for the language `#1'. Using the pattern for}%
\typeout{** the default language instead.}%
\else
\language=\csname l@#1\endcsname
\fi
#2}}
\providecommand{\BIBdecl}{\relax}
\BIBdecl

\bibitem{chertkov_cascading_2015}
M.~Chertkov, S.~Backhaus, and V.~Lebedev, ``Cascading of fluctuations in
  interdependent energy infrastructures: {Gas}-grid coupling,'' \emph{Applied
  Energy}, vol. 160, pp. 541--551, 2015.

\bibitem{zlotnik_coordinated_2016}
A.~Zlotnik, L.~Roald, S.~Backhaus, M.~Chertkov, and G.~Andersson, ``Coordinated
  scheduling for interdependent electric power and natural gas
  infrastructures,'' \emph{IEEE Transactions on Power Systems}, vol.~32, no.~1,
  pp. 600--610, 2016, publisher: IEEE.

\bibitem{byeon2020awareness}
G.~Byeon and P.~Van~Hentenryck, ``Unit commitment with gas network awareness,''
  \emph{IEEE Transactions on Power Systems}, vol.~35, no.~2, pp. 1327--1339,
  2020.

\bibitem{Bayani_2022}
R.~Bayani and S.~D. Manshadi, ``Natural gas short-term operation problem with
  dynamics: A rank minimization approach,'' \emph{{IEEE} Transactions on Smart
  Grid}, vol.~13, no.~4, pp. 2761--2773, jul 2022.

\bibitem{roald2020uncertainty}
L.~A. Roald, K.~Sundar, A.~Zlotnik, S.~Misra, and G.~Andersson, ``An
  uncertainty management framework for integrated gas-electric energy
  systems,'' \emph{Proceedings of the IEEE}, vol. 108, no.~9, pp. 1518--1540,
  2020.

\bibitem{eia2023}
\BIBentryALTinterwordspacing
``{U.S.} {E}nergy {I}nformation {A}dministration,'' 2023. [Online]. Available:
  \url{https://www.eia.gov/todayinenergy/detail.php?id=42776}
\BIBentrySTDinterwordspacing

\bibitem{leahy2012cost}
\BIBentryALTinterwordspacing
E.~Leahy, C.~Devitt, S.~Lyons, and R.~S. Tol, ``The cost of natural gas
  shortages in ireland,'' \emph{Energy Policy}, vol.~46, pp. 153--169, 2012.
  [Online]. Available:
  \url{https://www.sciencedirect.com/science/article/pii/S0301421512002522}
\BIBentrySTDinterwordspacing

\bibitem{kim2015estimation}
\BIBentryALTinterwordspacing
K.~Kim, H.~Nam, and Y.~Cho, ``Estimation of the inconvenience cost of a rolling
  blackout in the residential sector: The case of south korea,'' \emph{Energy
  Policy}, vol.~76, pp. 76--86, 2015. [Online]. Available:
  \url{https://www.sciencedirect.com/science/article/pii/S0301421514005643}
\BIBentrySTDinterwordspacing

\bibitem{abu-khalifa2023}
Y.~A. Abu-Khalifa and A.~B. Birchfield, ``Techniques for creating synthetic
  combined electric and natural gas transmission grids,'' \emph{IEEE
  Transactions on Industry Applications}, vol.~59, no.~4, pp. 4734--4743, 2023.

\bibitem{hyett2023control}
\BIBentryALTinterwordspacing
C.~Hyett, L.~Pagnier, J.~Alisse, L.~Sabban, I.~Goldshtein, and M.~Chertkov,
  ``Control of line pack in natural gas system: Balancing limited resources
  under uncertainty,'' \emph{Proceeding of Pipeline Simulation Interest Group
  (PSIG) 2023 meeting}, 2023. [Online]. Available:
  \url{https://arxiv.org/abs/2304.01955}
\BIBentrySTDinterwordspacing

\bibitem{CEC2019}
\BIBentryALTinterwordspacing
P.~Deaver, M.~Kootstra, and R.~MacDonald, ``Updating thermal power plant
  efficiency measures and operational characteristics for production cost
  modeling,'' California Energy Commission, Staff Report CEC-200-2019-001,
  January 2019. [Online]. Available:
  \url{https://www.energy.ca.gov/sites/default/files/2021-06/CEC-200-2019-001.pdf}
\BIBentrySTDinterwordspacing

\bibitem{sutton2018reinforcement}
R.~S. Sutton and A.~G. Barto, \emph{Reinforcement Learning: An Introduction},
  2nd~ed.\hskip 1em plus 0.5em minus 0.4em\relax MIT Press, 2018, originally
  published in 1998.

\end{thebibliography}

\end{document}